\documentclass[prd,preprint,tightenlines]{revtex4}

\usepackage{graphicx}

\begin{document}

\title{Antisymmetric rank--2 tensor unparticle physics}
\author{Tae-il Hur$^{ab}$, P. Ko$^b$ and Xiao-Hong Wu$^{b}$}
\affiliation{
$^a$ Department of Physics, KAIST, Daejeon 305-701, Korea \\
$^b$ School of Physics, Korea Institute for Advanced Study,
         207-43, Cheongryangri 2-dong, Dongdaemun-gu, Seoul 130-722, Korea}

\begin{abstract}
We present the phenomenology of antisymmetric rank-2 tensor unparticle
operator ${\cal O}_{\cal U,A}^{\mu\nu}$ with scaling dimension $d_{\cal U}$.
We consider the physical effects of operator 
$O_{\cal U,A}^{\mu\nu}$
in $Z^0$ boson invisible decays $Z^0 \to {\cal U}$,
$Z^0 \to b \bar{b}$ channel,
the electroweak precision observable $S$ parameter,
and the muon anomalous magnetic dipole moment.
The $Z^0$ boson invisible decay gives a very stringent
constraint in the $( \Lambda_{\cal U} , M_{\cal U})$ plane,
and only small $r \equiv \Lambda_{\cal U} / M_{\cal U} \lesssim 0.1$  
is favored, when $\Lambda_{\cal U}$ is order of several $100$ GeV.
When the phenomenological parameter $\mu$,
which parameterizes the scale invariance breaking,
goes to $0$, the $S$ parameter and the muon $(g-2)$ diverge
for $1 < d_{\cal U} < 2$,
while for non-zero $\mu$, there will be constraints
on $(\Lambda_{\cal U} , M_{\cal U})$ which are more stringent than those
obtained from collider experiments.
\end{abstract}

\maketitle

\section{Introduction}

Recently Georgi proposed an interesting possibility that there might be
a hidden sector with operators $O_{UV}$'s that flows into scale invariant 
theory at low energy scale $\Lambda_{\cal U}$ with operators $O_{\cal U}$'s. 
The hidden sector operators $O_{UV}$'s in the UV theory can interact 
with the SM sector by nonrenormalizable interactions that are generated 
at some scale $M_{\cal U}$,
which match to the scale
invariant interaction with the Standard Model (SM) sector
below $\Lambda_{\cal U}$    
\cite{Georgi:2007ek,Georgi:2007si}. 
Schematically, one has the following picture: 
\begin{equation}
C_n { {\cal O}_{SM} {\cal O}_{UV} \over M_U^{d_{UV} + n - 4} }
\longrightarrow 
C_n^i {\Lambda_{\cal U}^{d_{UV} - d_{\cal U}} \over M_{U}^{d_{UV} + n - 4}}
{\cal O}_{SM,i}^n {\cal O}_{\cal U} 
 \equiv 
{ C_n^i \over \Lambda_{n}^{d_{\cal U} + n - 4}}
{\cal O}_{SM,i}^n {\cal O}_{\cal U}
\end{equation}
where $d_{UV}$, $d_{\cal U}$ and $n$ are 
the scaling dimensions of the UV operator in the Banks-Zaks (BZ) sector 
\cite{Banks:1981nn}, the unparticle operator $O_{\cal U}$,
and the SM operator of a type $i$, ${\cal O}_{SM,i}^n$. 
Triggered by this intriguing suggestion, a lot of phenomenological
analysis have been  done for both low and high energy processes involving 
unparticle operators 
\cite{Cheung:2007ue,Luo:2007bq,Chen:2007vv,Ding:2007bm,Liao:2007bx,
Aliev:2007qw,Li:2007by,Duraisamy:2007aw,Lu:2007mx,Greiner:2007hr,
Stephanov:2007ry,Fox:2007sy,Davoudiasl:2007jr,Choudhury:2007js,Chen:2007qr,
Aliev:2007gr,Mathews:2007hr,Zhou:2007zq,Ding:2007zw,Chen:2007je,Liao:2007ic,
Bander:2007nd,Rizzo:2007xr,Cheung:2007ap,Goldberg:2007tt,Chen:2007zy,
Zwicky:2007vv, Kikuchi:2007qd, Mohanta:2007ad,
Huang:2007ax, Krasnikov:2007fs, Lenz:2007nj, van der Bij:2007um,
Choudhury:2007cq, Zhang:2007ih, Li:2007kj, Nakayama:2007qu,
Deshpande:2007jy, Ryttov:2007sr, Mohanta:2007uu,
Delgado:2007dx, Cacciapaglia:2007jq, Neubert:2007kh,
Luo:2007me, Hannestad:2007ys, Deshpande:2007mf,
Das:2007nu, Bhattacharyya:2007pi, Liao:2007fv,
Majumdar:2007mp, Alan:2007ss, Freitas:2007ip,
Gogoladze:2007jn, Chen:2007pu}.
So far, most works have considered scalar and vector unparticle operators,  
and a few works on the rank--2 symmetric tensors, especially its
modification of the Newtonian gravity potential.

In this letter, we study the antisymmetric rank--2 tensor
unparticle operator,  ${\cal O}_{\cal U,A}^{\mu\nu}$:
its spectral representation, propagator and phenomenological implications.
This operator has a unique property that it can have 
coupling with the dimension--2 SM operator $B_{\mu\nu}$, the field strength
tensor of $U(1)_Y$ gauge boson.
Therefore it can contribute to the invisible decay width of $Z^0$ boson.
Also it can modify $Z\rightarrow b\bar{b}$,
the Peskin-Takeuchi $S$ parameter, and the muon $(g-2)$,
when we simultaneously consider its interactions with other SM operators. 
In the following, we consider these observables in the presence of
antisymmetric rank--2 tensor unparticle operator.

Before closing this section, we would like to define the conventions 
and the normalizations  of interactions between the unparticle operators 
and the SM. In most papers on unparticle phenomenology, 
the scale $\Lambda_n$ in Eq.~(1) is widely used, and bounds on them are 
derived assuming the coupling $C_{n}^i = O(1)$.  
However these scales $\Lambda_{n}$'s are derived from 
the first line of Eq.~(1), and it depends on the dimension of the SM 
operator that couples to, and in general $\Lambda_3 \neq \Lambda_4$. 
Scales $\Lambda_n$'s can be expressed in terms of fundamental scale 
$M_U$ and $\Lambda_{\cal U}$, and the scaling dimensions
$d_{\rm UV}$, $d_{\cal U}$ and $n$.
The expressions for $\Lambda_{i=2,3,4}$ 
can be found in Ref.~\cite{Bander:2007nd}:
\[
\Lambda_2 = r^{\frac{d_{\rm UV} -2}{2-d_{\cal U}}} \Lambda_{\cal U},
~~~~~
\Lambda_3 = \left(\frac{1}{r}\right)^{\frac{d_{\rm UV}-1}{d_{\cal U}-1}}
 \Lambda_{\cal U},
~~~~~
\Lambda_4 = \left(\frac{1}{r}\right)^{\frac{d_{\rm UV}}{d_{\cal U}}}
 \Lambda_{\cal U} ,
\]
where $r \equiv \Lambda_{\cal U} / M_{\cal U}$ is less or equal to $1$
by definition. For $1 < d_{\cal U} < 2 < d_{UV}$, one has
\begin{equation}
\Lambda_{2} < M_U < \Lambda_{4} < \Lambda_3 . 
\end{equation}
%
%
In this analysis, we will follow the conventions of
Ref.~\cite{Bander:2007nd} for the scales $\Lambda_n$'s and the couplings 
$C_n^i$'s and study various observables that could be affected by the 
antisymmetric rank--2 tensor unparticle operator $O_{\cal U,A}^{\mu\nu}$. 
Then the bounds on the scales $\Lambda_n$'s will be cast into those on 
$( \Lambda_{\cal U} , M_{\cal U} )$ for $d_{UV} = 3$. This will facilitate 
the comparison with the existing bounds in much clearer way. 

\section{two-point function and propagator}
The antisymmetric tensor $O_{\mathcal U,A}^{\mu\nu} =
 - O_{\mathcal U,A}^{\nu\mu}$
can be decomposed into magnetic and electric vectors,
with the index $\mu\nu=ij$, $i,j=1,2,3$ for magnetic components, 
and $\mu\nu=0i$ for electric components~\cite{VanNieuwenhuizen:1973fi}.
Using the scale symmetry, the two-point function for
$O_{\mathcal U,A}^{\mu\nu}$ can be written as
\begin{eqnarray}
\label{eq:2pt}
\left\langle 0 | O_{\mathcal U,A}^{\mu\nu}(x) 
O_{\mathcal U,A}^{\rho\sigma}(0) |
  0 \right\rangle
 &=& \int \frac{d^4 p}{(2\pi)^4} e^{-i p x} A_{d_{\mathcal U}}
  \theta(p^0) \theta(p^2)
  \Pi^{\mu\nu\rho\sigma} ( p^2 )^{d_{\mathcal U} -2} ,
\end{eqnarray}
where the projection operator $\Pi^{\mu\nu\rho\sigma}$ is defined as follows
~\cite{VanNieuwenhuizen:1973fi}:
\[
\Pi^{\mu\nu\rho\sigma} = \frac{1}{2}
  (P^{\mu\rho} P^{\nu\sigma} - P^{\mu\sigma} P^{\nu\rho})
  \]
for magnetic component, and
\[
\Pi^{\mu\nu\rho\sigma} = \frac{1}{2}
  (P^{\mu\rho} \omega^{\nu\sigma} - P^{\mu\sigma} \omega^{\nu\rho}
   - P^{\nu\rho} \omega^{\mu\sigma} + P^{\nu\sigma} \omega^{\mu\rho} )
\]
for electric part, with
\[
P^{\mu\nu} = g^{\mu\nu} - p^\mu p^\nu p^{-2} , \ \ \
\omega^{\mu\nu} = p^\mu p^\nu p^{-2} .
\]
We use the same normalization for the overall factor $A_{d_{\mathcal U}}$
as Georgi \cite{Georgi:2007ek,Georgi:2007si}:
\[
A_{d_{\mathcal U}} = {16 \pi^{5/2} \over ( 2 \pi )^{2d} } ~{\Gamma ( d + 1/2 )
\over \Gamma ( d -1 ) \Gamma ( 2 d )} .
\]
It is straightforward to derive the propagator using the above expression
for the two-point function:
\begin{eqnarray}
\left\langle 0 | T ( O_{\mathcal U,A}^{\mu\nu}(x)
  O_{\mathcal U,A}^{\rho\sigma}(0) ) | 0 \right\rangle
 &=& i \frac{A_{d_{\mathcal U}}}{2}
  \int \frac{d^4 p}{(2\pi)^4} e^{-i p x}
  \frac{\Pi^{\mu\nu\rho\sigma}}{\sin(d_{\mathcal U}\pi)}
  (- p^2 - i \epsilon)^{d_{\mathcal U} -2}
\label{eq:prop}
\end{eqnarray}
These two equations (\ref{eq:2pt}) and (\ref{eq:prop}) are 
one of the main results of this paper, and necessary ingredient 
in order that we calculate physical observables in the following.

In addition to ${\cal O}_{\cal U,A}^{\mu\nu}$,
there is one more unparticle operator
that can couple to the dimension--2 SM operator $H^\dagger H$
\cite{Fox:2007sy}:
\[
{c_2 \over \Lambda_2^{d_U -2}}~{\mathcal O_S} H^\dagger H .
\]
After the EWSB, this operator will generate a mass scale $\mu$, and thus
induces conformal symmetry breaking in the unparticle sector. 
Bander et al. \cite{Bander:2007nd} argue that 
\[
\left( { \mu \over \Lambda_4 } \right) \lesssim \left( 10^{-3} 
\right)^{1/d_{\cal U}} .
\]
%
Following Ref.~\cite{Fox:2007sy}, we parameterize this conformal symmetry
breaking by phenomenological effective mass scale $\mu$, and modify
the two-point function and propagator  as follows:
\begin{eqnarray}
\left\langle 0 | O_{\mathcal U,A}^{\mu\nu}(x) O_{\mathcal U,A}^{\rho\sigma}(0) |
  0 \right\rangle
 &=& \int \frac{d^4 p}{(2\pi)^4} e^{-i p x} A_{d_{\mathcal U}}
  \theta(p^0) \theta(p^2 - \mu^2)
  \Pi^{\mu\nu\rho\sigma} ( p^2 - \mu^2 )^{d_{\mathcal U} -2} \\
\left\langle 0 | T ( O_{\mathcal U,A}^{\mu\nu}(x)
  O_{\mathcal U,A}^{\rho\sigma}(0) ) | 0 \right\rangle
 &=& i \frac{A_{d_{\mathcal U}}}{2}
  \int \frac{d^4 p}{(2\pi)^4} e^{-i p x}
  \frac{\Pi^{\mu\nu\rho\sigma}}{\sin(d_{\mathcal U}\pi)}
  [- (p^2 - \mu^2) - i \epsilon ]^{d_{\mathcal U} - 2}
\end{eqnarray}
In the following, we find that the contributions of an antisymmetric 
rank-2 tensor unparticle operator to the $S$ parameter and the muon $(g-2)$ 
are proportional to $( \mu^2 )^{d_{\cal U}-2}$, 
so that they are divergent for $\mu=0$ and $1 < d_{\cal U} <2$. 
Therefore we have to keep the nonzero $\mu$ scale for some observables 
considered in this work.

\section{Phenomenology}

In this section, we consider the physical effects of antisymmetric
unparticle operator $O_{\mathcal U,A}^{\mu\nu}$ of scaling
dimension $d_{\mathcal U}$ on 
the invisible decay width of $Z^0$ boson  ($Z \to {\cal U}$), 
$R_b$ and $A_{\rm FB}^b$ in the $Z \to b \bar{b}$ decay channel, 
an electroweak precision observable $S$ parameter, 
and the muon anomalous magnetic dipole moment.
The relevant interaction terms involving the antisymmetric rank-2 tensor 
unparticle operator $O_{\mathcal U,A}^{\mu\nu}$ can be written as
\begin{eqnarray}
\label{lagrangian} {\mathcal L}_{\rm int} & = & \lambda_b
\frac{g^{'}}{\Lambda_2^{d_{\mathcal U}-2}}
  B_{\mu\nu} O_{\mathcal U,A}^{\mu\nu}
+ \widetilde{\lambda_b} \frac{g^{'}}{\Lambda_2^{d_{\mathcal U}-2}}
  \widetilde{B_{\mu\nu}} O_{\mathcal U,A}^{\mu\nu}
\nonumber \\
& + & \lambda_w \frac{g}{\Lambda_4^{d_{\mathcal U}}}
  (H^\dagger \tau^a H) W^a_{\mu\nu} O_{\mathcal U,A}^{\mu\nu}
+ \lambda_f \frac{y_f}{\Lambda_4^{d_{\mathcal U}}}
  \bar{f}_L H \sigma_{\mu\nu} f_R O_{\mathcal U,A}^{\mu\nu} \, .
\end{eqnarray}
Here the field strength tensor $B_{\mu\nu}$ of the $U(1)_Y$ gauge boson 
and its dual $\widetilde{B_{\mu\nu}}$ are gauge invariant dimension--2
operators, and the other two SM operators are of dimension 4.
We included the gauge couplings for gauge fields, and the Yukawa
couplings for Higgs-fermion couplings, and assumed that the
couplings $\lambda_i$'s are all order $O(1)$. This is in accord with 
Ref.~\cite{Bander:2007nd} and we can compare directly our results on 
$( \Lambda_{\cal U} , M_{\cal U} )$ with their results on scalar and vector 
unparticle operators derived from LEP/SLC.  

Note that the first two terms are unique to the antisymmetric rank--2 
tensor unparticle operator $O_{\mathcal U,A}^{\mu\nu}$ we consider in
this work. As mentioned in the previous section, there are only
two operators $O_{\mathcal U,A}^{\mu\nu}$ and the scalar
unparticle operator ${\cal O_{U,S}}$ that can couple to the
dimension--2 SM operators, which are $B_{\mu\nu}$ and $H^\dagger
H$, respectively.

\subsection{$Z^0  \to {\cal U}$}

In the presence of antisymmetric rank--2 tensor unparticle
operator, $Z^0$ boson can  decay to an invisible unparticle, 
$Z^0 \to {\cal U}$ through (i) the 1st term alone  and
(ii) the second order effects involving the 1st and the 3rd terms 
in Eq.~(\ref{lagrangian}).  The 2nd term of Eq.(\ref{lagrangian}) does not 
give a nonvanishing result due to the Levi-Civita tensor.
Using the results obtained in Sec.~II, we can calculate the
$Z^0 \rightarrow {\mathcal U}$ decay width easily:
\begin{eqnarray}
\label{ZtoU}
\Gamma(Z^0 \to U) &=& 2 g^{\prime 2} s_w^2 m_Z A_{d_{\mathcal U}} \lambda_b^2
  \left( \frac{m_Z^2 - \mu^2}{\Lambda_2^2} \right)^{d_{\mathcal U} - 2}
  \left(1 + \frac{c_w}{4 s_w} \frac{\lambda_w}{\lambda_b}
   \frac{v^2}{\Lambda_4^2}
   \frac{\Lambda_2^{d_{\mathcal U} - 2}}{\Lambda_4^{d_{\mathcal U} - 2}}
   \right)^2
  \theta(m_Z^2 - \mu^2)
\end{eqnarray}
where $s_w \equiv \sin^2 \theta_W$ and $c_w \equiv \cos \theta_W$. 
We note that in the $Z$ rest frame, only the electric component of
antisymmetric tensor contributes to the invisible 
$Z\rightarrow {\mathcal U}$ decay.
The contributions of a vector unparticle operator 
$\partial^\mu O_V^\nu$ coupled to dimension--2 SM operator $B_{\mu\nu}$ 
have already been considered in ref.~\cite{Chen:2007zy}. 
However the operator we considered here, the  term with dimension--2 SM 
operator in Eq.~(\ref{lagrangian})  gives the dominant contribution.

The $Z^0$ boson properties have been well studied, and its invisible decay
width is quite consistent with the SM predictions with three light neutrinos:
\[
\Gamma^{\rm invis}_{\rm exp} ( Z^0 ) = ( 499.0 \pm 1.5 ) ~{\rm MeV} ,
\ {\rm vs.} \
\Gamma_{\rm SM} ( Z^0 \rightarrow \nu \bar{\nu}) =
( 501.65 \pm 0.11 )~{\rm MeV} .
\]
This leaves  little room to an additional invisible decay width of
$Z^0$ boson, for example, into the unparticle sector.
We assume that the room for the invisible $Z^0 \to {\cal U}$ decay 
is the uncertainty of experimental uncertainty in 
$\Gamma^{\rm invis}$, namely $1.5$ MeV.

\begin{figure}
\centerline{\includegraphics[width=8cm] {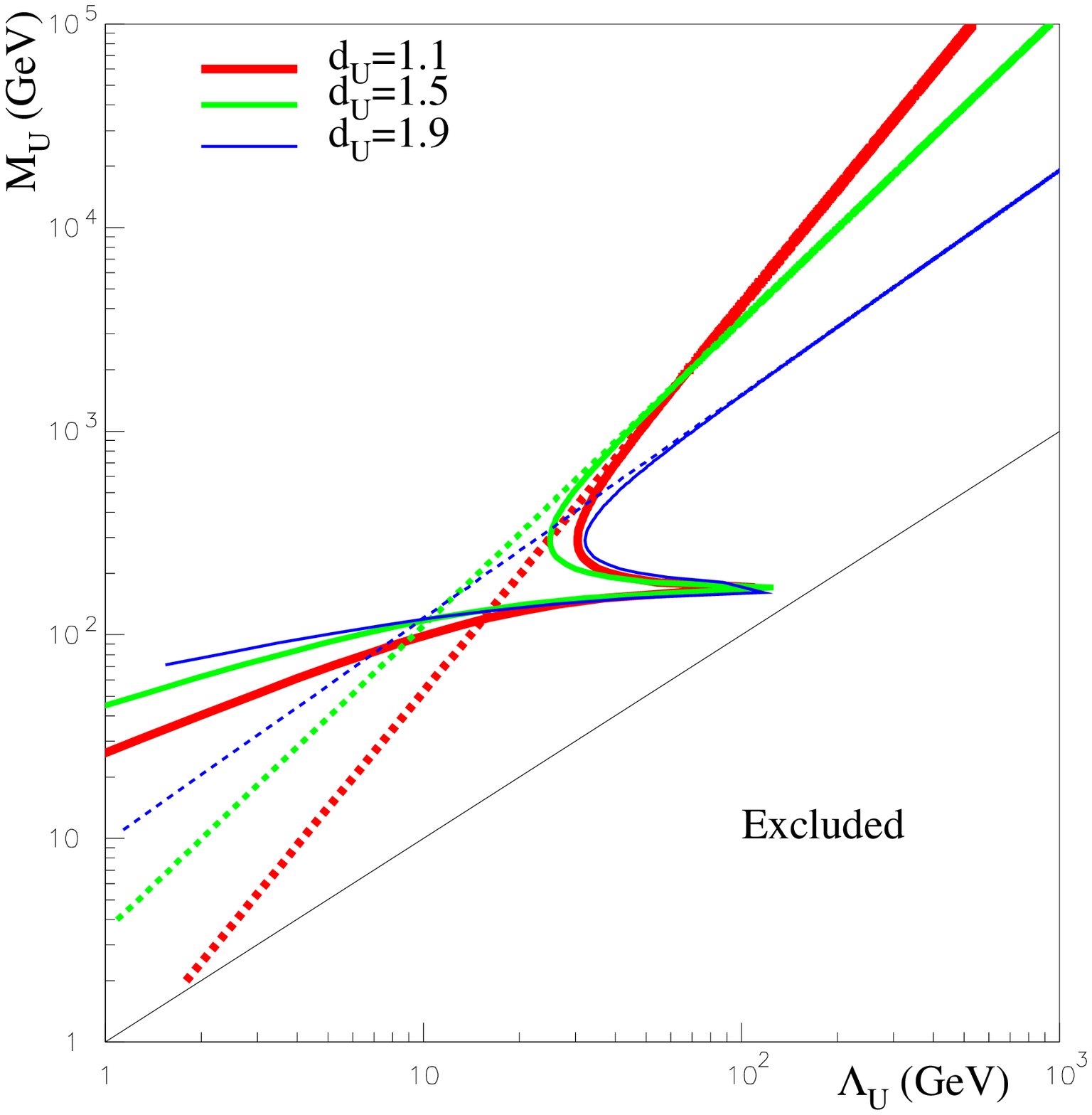}
\includegraphics[width=8cm] {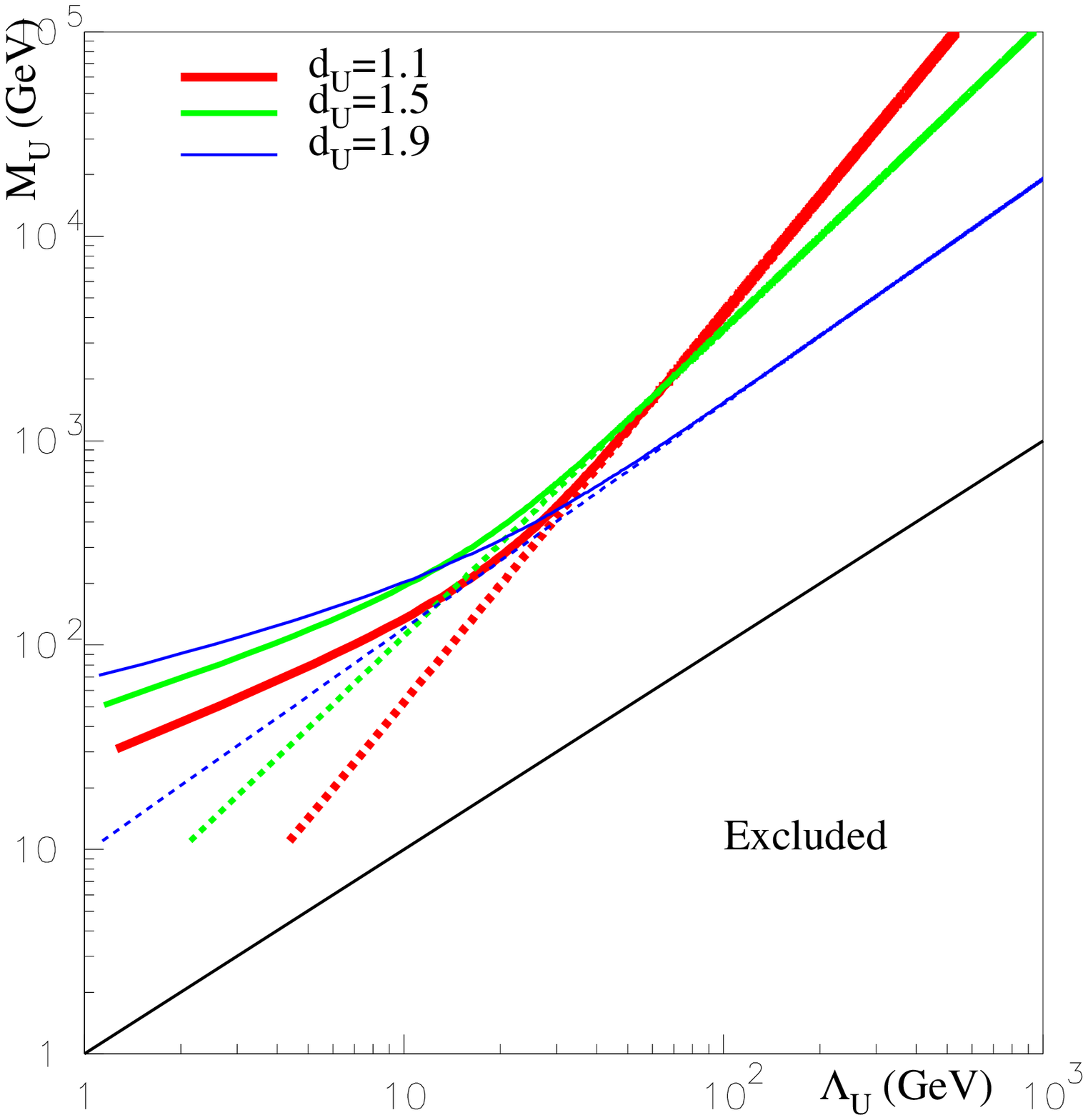}}
\vspace{-8mm}
\centerline{(a) \hspace{7cm} (b)}
\centerline{\includegraphics[width=8cm] {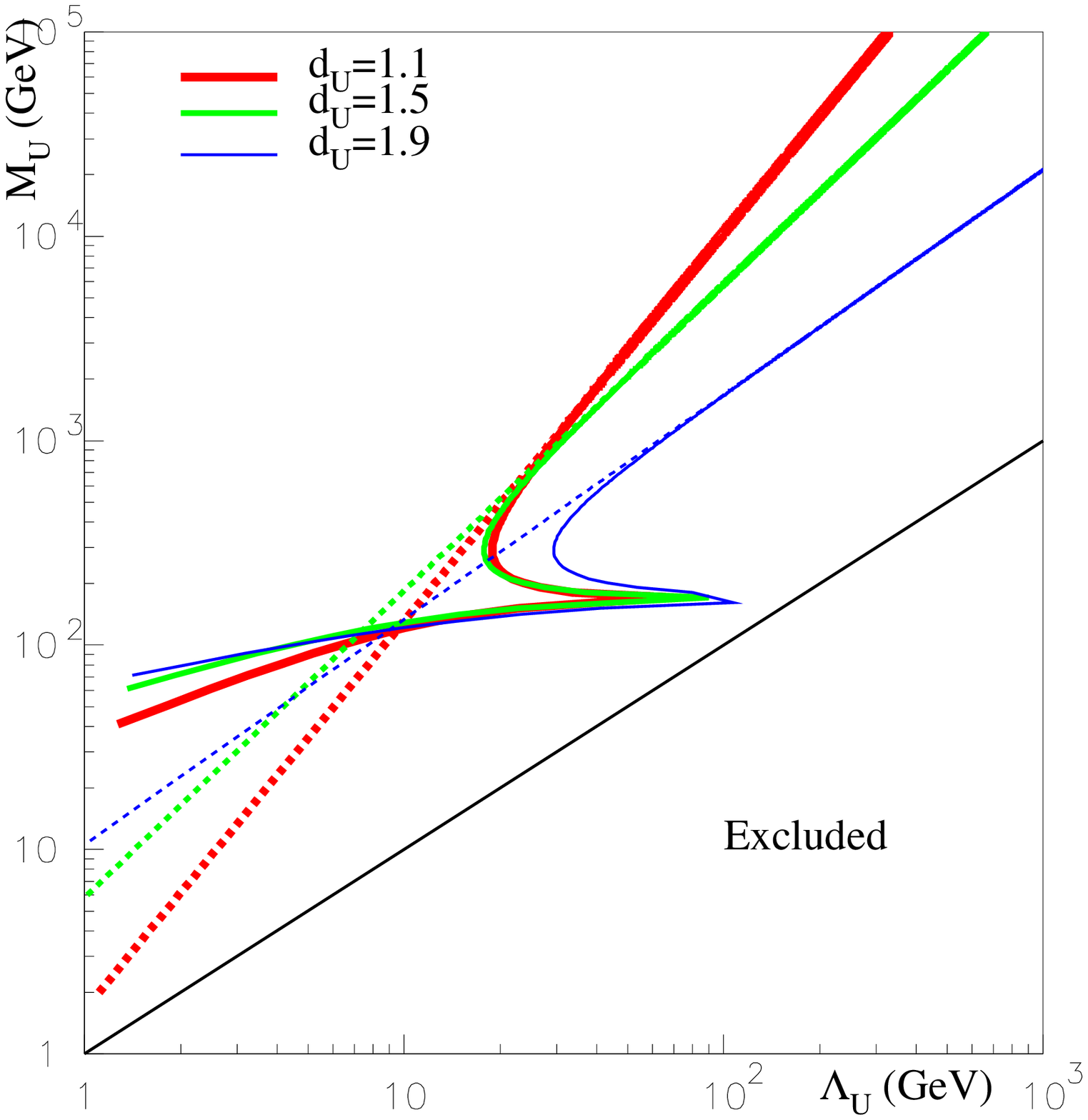}
\includegraphics[width=8cm] {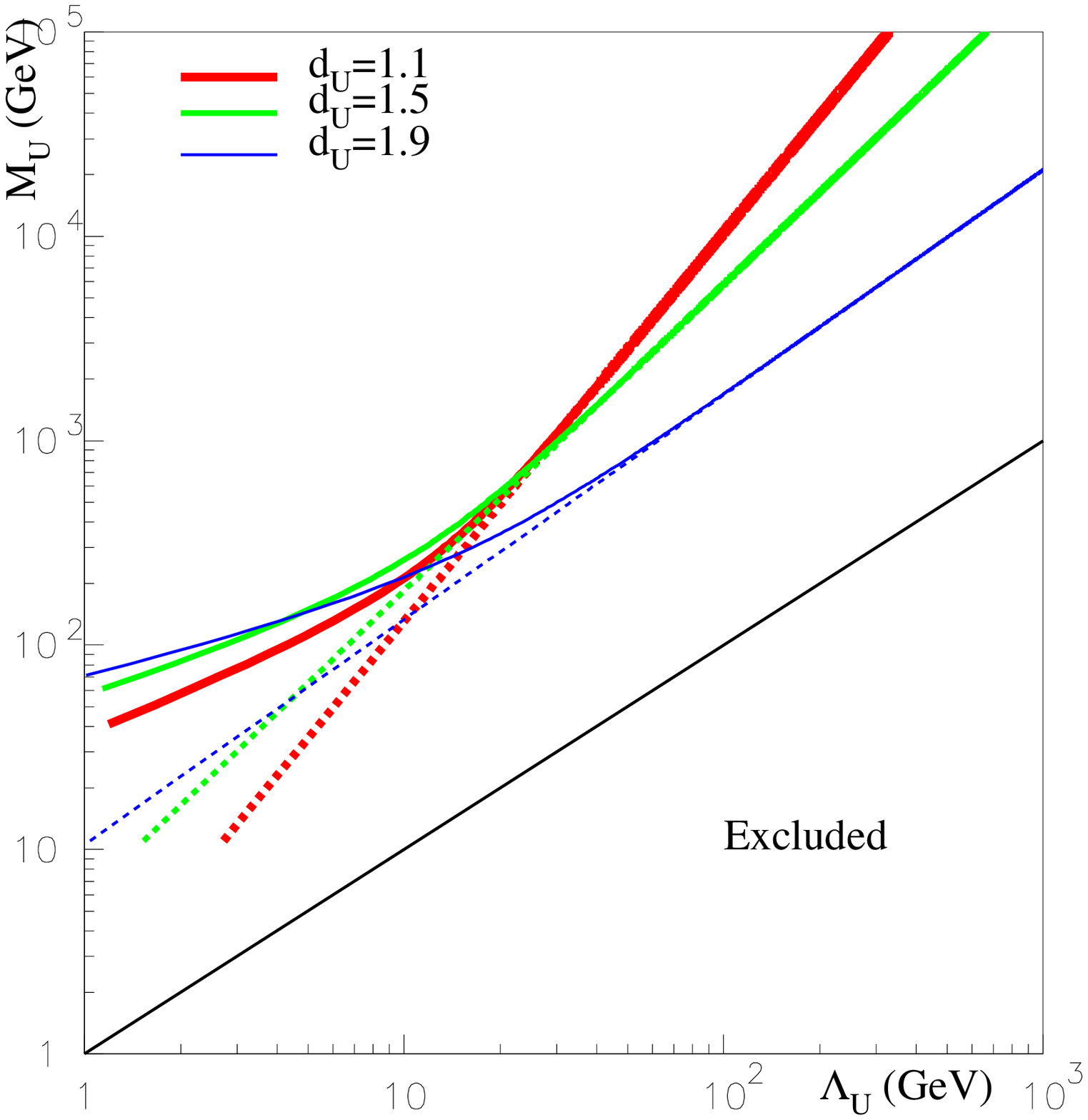}}
\vspace{-8mm}
\centerline{(c) \hspace{7cm} (d)}
\caption{ \label{invis}
The contour plot of $\Gamma(Z\to U)$ with $\Gamma(Z\to U) = 1.5$MeV
in $\Lambda_{\cal U}$ and $M_{\cal U}$ plane
for $d_{\cal U}=1.1$ (red), $d_{\cal U}=1.5$ (green),
and $d_{\cal U}=1.9$ (blue).
(a) and (b) correspond to $\mu=0$,
(c) and (d) correspond to $\mu=85$ GeV.
$\lambda_b=-\lambda_w=1$ in case (a) and (c),
$\lambda_b=\lambda_w=1$ in case (b) and (d).
In the region above the corresponding curve for different $d_{\cal U}$,
$\Gamma(Z\to U)$ is smaller than $1.5$MeV.
The region below $\Lambda_{\cal U}=M_{\cal U}$ (black) is excluded.
}
\end{figure}

In the numerical analyses of $Z^0 \rightarrow {\cal U}$,  
we fix the dimensionless coupling to $1$,
$d_{\rm UV}=3$ for $d_{\cal U}=1.1$, $1.5$, $1.9$,
and choose two different $\mu=0$ and  $\mu = 85$ GeV.
We note that the variation of $\mu$ parameter with $\mu < m_Z$  does not
affect the results very much, as $\mu$ enters Eq.~(\ref{ZtoU})
through the combination of $( m_Z^2 - \mu^2 )^{d_{\cal U} - 2}$.
The contour plots for $\Gamma(Z^0 \to {\cal U})$
in the $ ( \Lambda_{\cal U} , M_{\cal U} )$ plane
is shown in Fig.~\ref{invis}. 
The dash lines in the figure correspond to
the contribution from the first term in Eq.~(\ref{lagrangian}) only,
which is the contribution in Eq.~(\ref{ZtoU})
without the second term $\frac{c_w}{4s_w}...$ in the squared bracket.
We find that the contribution from second term $\frac{c_w}{4s_w}...$
is not small when $\Lambda_{\cal U}$ is below several tens of GeV,
while for larger $\Lambda_{\cal U}$,
the contribution is dominated by the first term,
as we can see that the dash and solid line overlap.
This is because the second term
$\frac{c_w}{4 s_w} \frac{\lambda_w}{\lambda_b}
   \frac{v^2}{\Lambda_4^2}
   \frac{\Lambda_2^{d_{\mathcal U} - 2}}{\Lambda_4^{d_{\mathcal U} - 2}}
 = \frac{c_w}{4 s_w} \frac{\lambda_w}{\lambda_b}
  \frac{v^2}{\Lambda_{\cal U}^2} r^2 \sim \frac{v^2}{\Lambda_{\cal U}^2}$
with $r = \frac{\Lambda_{\cal U}}{M_{\cal U}} \le 1$,
when we replace $\Lambda_{2,4}$ with $M_{\cal U}$ and $\Lambda_{\cal U}$.
That factor is suppressed when $\Lambda_{\cal U}$ is larger than 
the electroweak scale $v$. 
Hence, the contribution from the third term of Eq.~(\ref{lagrangian})
is suppressed compared with the one from the first term.
We also change the sign of dimensionless parameter
$\lambda_w$ to $-1$ in case (a) and (c) to show
the contribution of the second term of Eq.~(\ref{ZtoU}),
because $\lambda_w$ only enter the second term in the formula.
For larger $\Lambda_{\cal U}$ of order $100$ GeV,
the second term is suppressed,
the curve in case (a) and (c) is similar to (b) and (d)
in the large $\Lambda_{\cal U}$ region.
We have similar results in the discussion of
other physical quantities later on.
We find that the constraint from the invisible decay of $Z$ is very 
stringent, only small $r = \Lambda_{\cal U} / M_{\cal U} \lesssim 0.1$ 
is favored.
We note that  $\Gamma(Z^0 \to {\cal U}) =0$ if $\mu \geq m_Z$,
due to the $\theta$ function in Eq.~(\ref{ZtoU}), and there would be no 
constraint from the invisible $Z^0$ decay width on the unparticle physics.
Comparing Fig.~\ref{invis} with the plots in Ref.~\cite{Bander:2007nd}, 
we can conclude that the constraint from the invisible $Z^0$ decay width 
is much more stringent than those from the effects of scalar and vector 
unparticle operators at LEP/SLC experiments as long as $\mu < m_Z$. 

\subsection{$Z ^0 \to \bar{b} b$}

The decay $Z^0 \rightarrow b\bar{b}$ is an important process not only
within the SM, but also in many models beyond the SM, since any new physics
scenarios which deals the 3rd generation quarks differently from the light
two families can affect this decay, and thus is strongly constrained by
electroweak precision data.
In the model independent effective theory approach, 
this channel has been discussed in \cite{Datta:1997qy} 
including the magnetic operators.
In the framework of unparticle physics,
vector unparticle contribution to $Z^0 \to \bar{b}b$ has been investigated
in ref.~\cite{Luo:2007me}.

\begin{figure}
\centerline{\includegraphics[width=8cm] {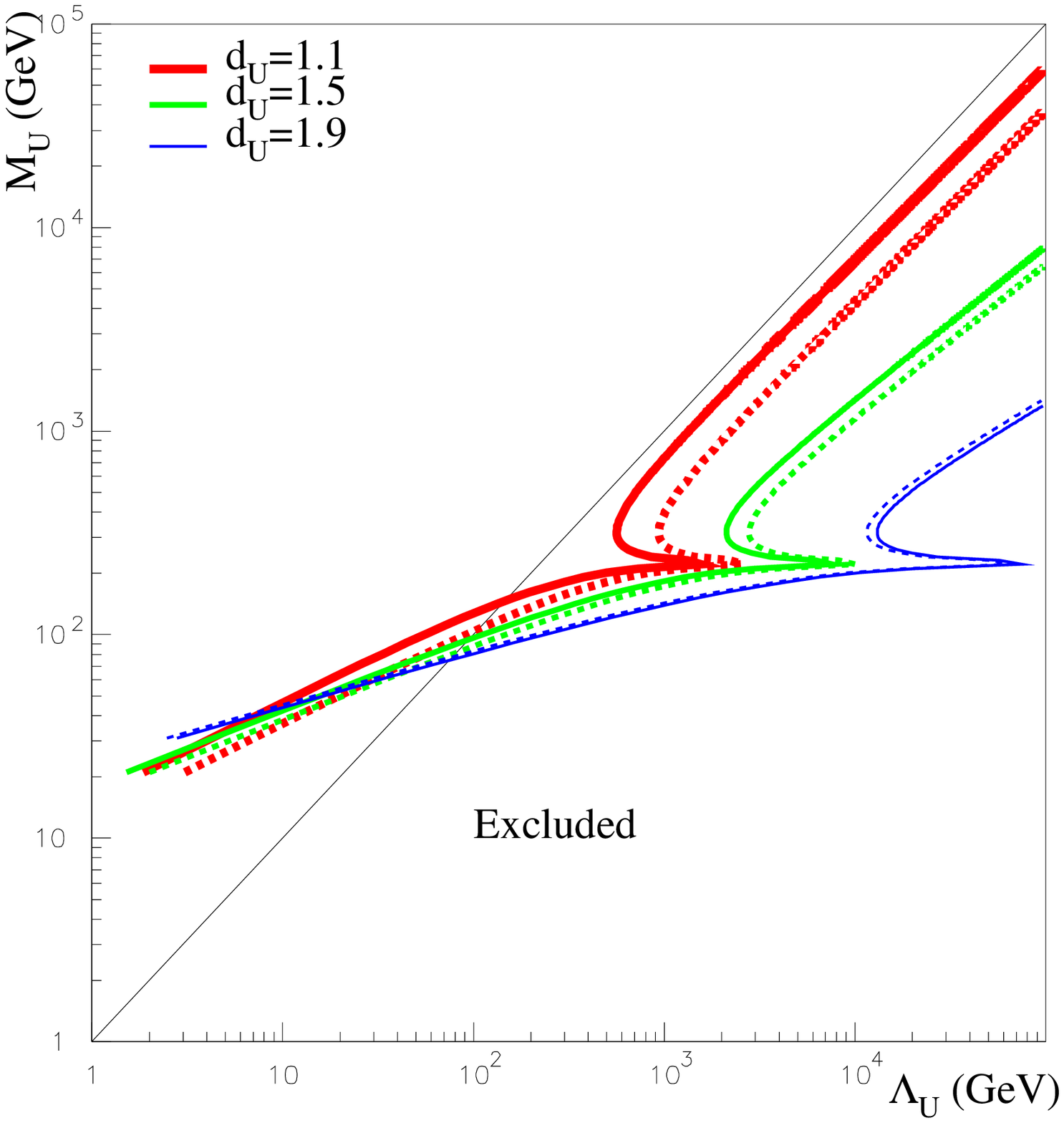}
\includegraphics[width=8cm] {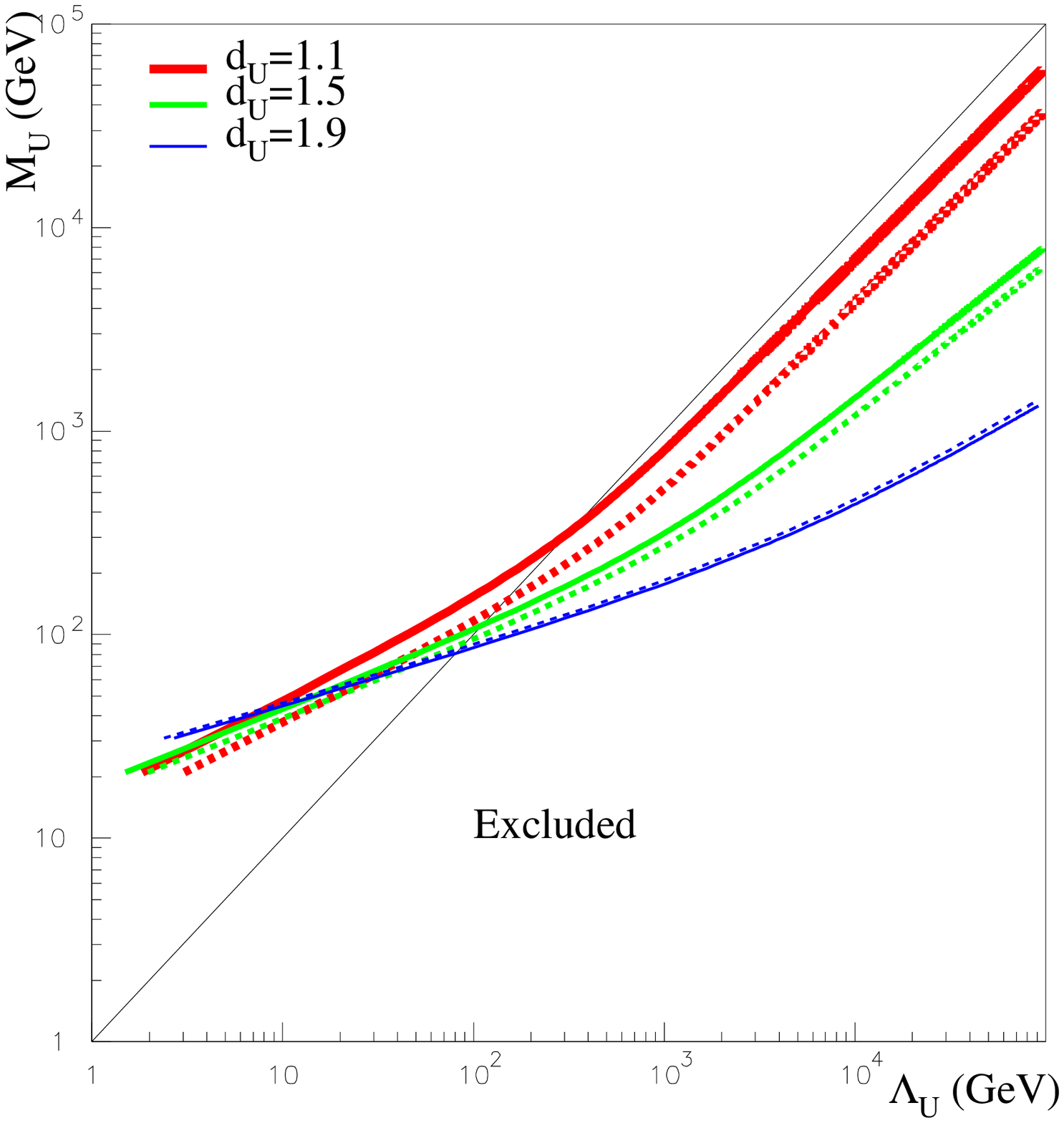}}
\vspace{-8mm}
$(a)$ \hspace{7.5cm} $(b)$
\caption{ \label{zbb}
The contour plot of $A^b_{\rm FB}$ (dash) and $R_b$ (solid)
within $1\sigma$ bound
in $\Lambda_{\cal U}$ and $M_{\cal U}$ plane
for $d_{\cal U}=1.1$ (red), $d_{\cal U}=1.5$ (green),
and $d_{\cal U}=1.9$ (blue) with $\mu=0$.
$\lambda_f=\lambda_b=-\lambda_w=1$ in case (a),
$\lambda_f=\lambda_b=\lambda_w=1$ in case (b).
The region above the corresponding curve for different $d_{\cal U}$
is allowed with $1\sigma$ bound.
}
\end{figure}

The antisymmetric rank--2 tensor unparticle operator
$O_{\mathcal U,A}^{\mu\nu}$ can contribute to the magnetic operator in the
decay $Z^0 \rightarrow b\bar{b}$  through $\lambda_b$ and $\lambda_f$ 
terms in Eq.~(\ref{lagrangian}).
We first calculate the coefficient $g^b_T$ of the magnetic operator defined
in \cite{Datta:1997qy}, and discuss the physical quantities
$A^b_{\rm FB}$ and $R_b$.
\begin{eqnarray}
\label{eq:zbb}
g^b_T = s_w^2 \lambda_b \lambda_f \frac{A_{d_{\mathcal U}}
   e^{-i(d_{\mathcal U} -2)\pi}}{2 \sin(d_{\mathcal U}\pi)}
  \frac{m_b m_z (m_z^2 - \mu^2)^{d_{\mathcal U} - 2}}
   {\Lambda_2^{d_{\mathcal U} - 2} \Lambda_4^{d_{\mathcal U}}}
  \left(1 + \frac{c_w}{4 s_w} \frac{g}{g^\prime} \frac{\lambda_w}{\lambda_b}
   \frac{v^2}{\Lambda_4^2}
   \frac{\Lambda_2^{d_{\mathcal U} - 2}}{\Lambda_4^{d_{\mathcal U} - 2}} \right)
\end{eqnarray}
Experimentally, both $A^b_{\rm FB}$ and $R_b$ have been
determined precisely, $A^b_{\rm FB} = 0.923 \pm 0.020$,
$R_b = 0.21629 \pm 0.00066$.
We fix the dimensionless parameters to $1$,
$d_{\rm UV}=3$ for $d_{\cal U}=1.1$, $1.5$, $1.9$,
and choose $\mu=0$.
The contour plot of $A^b_{\rm FB}$ and $R_b$ is presented
in Fig.~\ref{zbb}.
Most of the region above the $\Lambda_{\cal U} = M_{\cal U}$ (black)
line is allowed by the $1\sigma$ bound of both $A^b_{\rm FB}$ and $R_b$.
In that region, the unparticle effect is quit small and can be neglected.
We note that the variation of $\mu$ parameter
does not affect the results very much, as $\mu$ enters Eq.~(\ref{eq:zbb})
through the combination of $( m_Z^2 - \mu^2 )^{d_{\cal U} -2}$.

\subsection{$S$ parameter}

The Peskin--Takeuchi parameters, $S$, $T$ and $U$, have been introduced
in order to constrain new physics contributions to the gauge boson self
energy. 
The unparticle $O_{\mathcal U,A}^{\mu\nu}$ exchange through $\lambda_b$ and
$\lambda_w$ terms can induce the
dimension--6 operator $(H^\dagger \tau^a H) W^a_{\mu\nu} B^{\mu\nu}$,
which is directly related with the $S$ parameter:
\begin{eqnarray}
\label{sformulae}
S &=& 2 \frac{c_w}{s_w} g^2 g^{\prime 2} \lambda_b \lambda_w
  \frac{A_{d_{\cal U}}}{2 \sin(d_{\cal U}\pi)}
  \frac{v^2 (\mu^2)^{d_{\cal U} -2}}{\Lambda_2^{d_{\cal U} -2}
   \Lambda_4^{d_{\cal U}}} \,.
\end{eqnarray}
Let us note that the $S$ parameter is divergent if $\mu^2 = 0$ for 
$1< d_{\cal U} < 2$. This problem disappears if $d_{\cal U} > 2$,
but we do not consider this possibility here.

\begin{figure}
\centerline{\includegraphics[width=8cm] {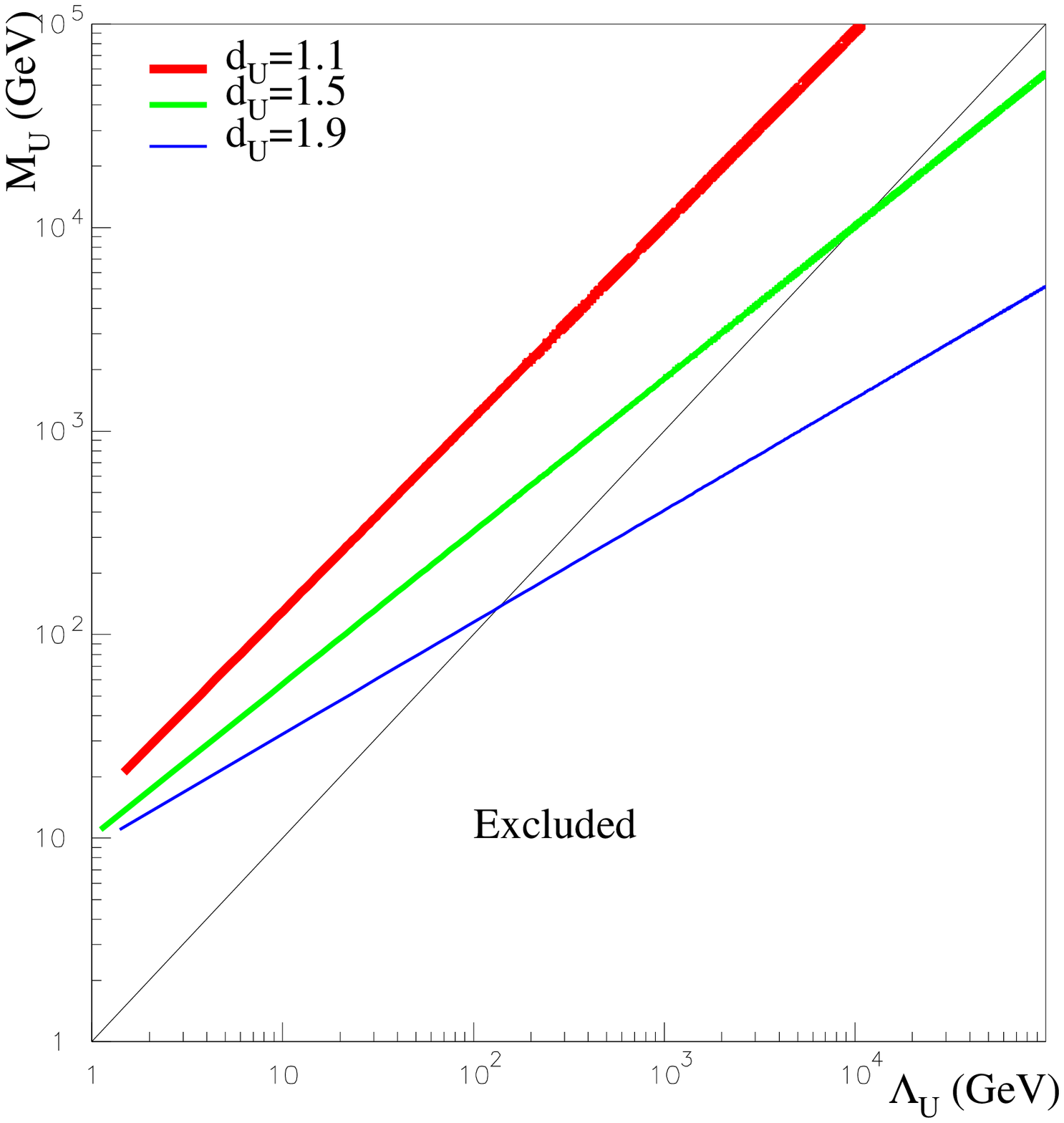}
\includegraphics[width=8cm] {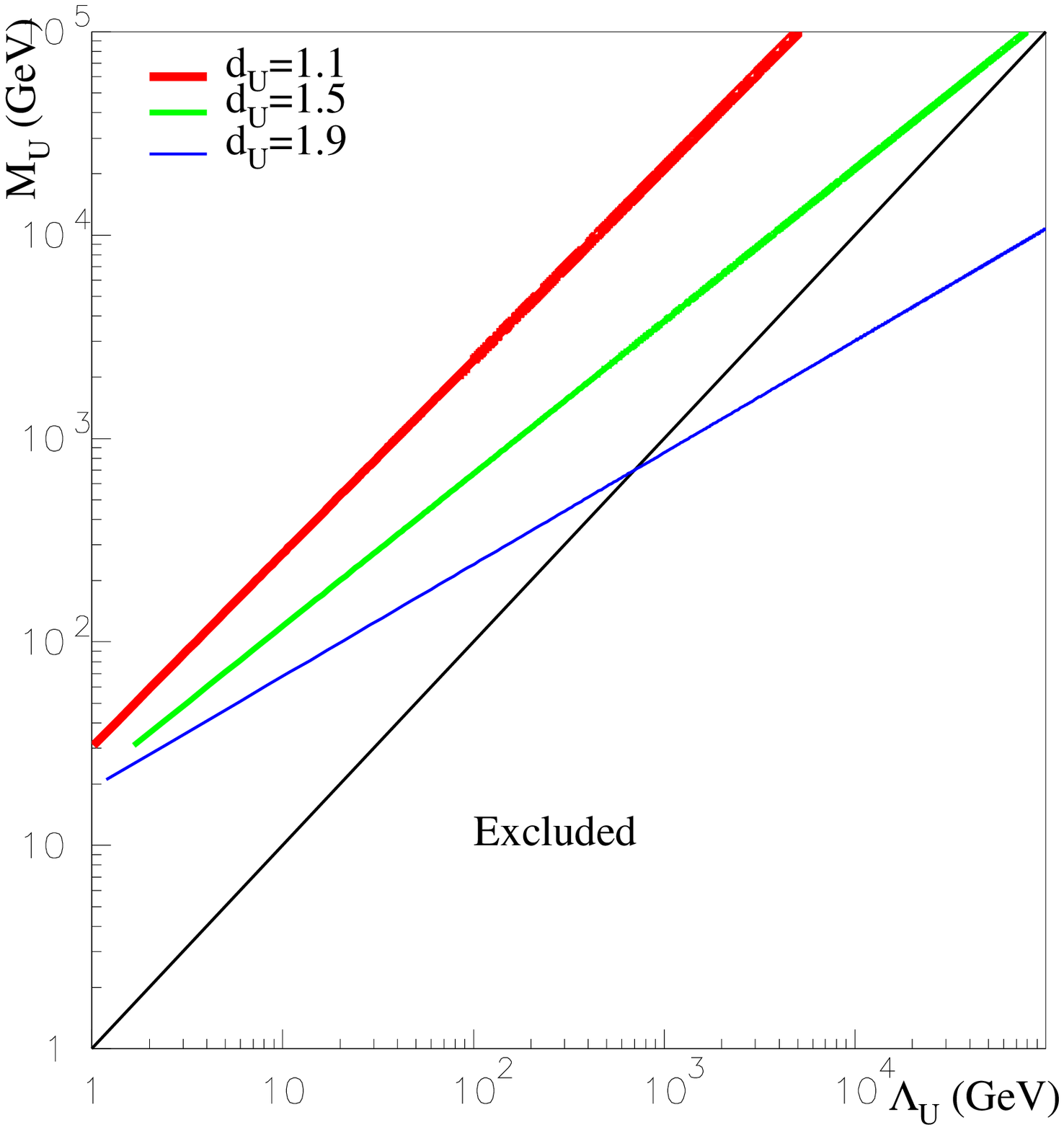}}
\vspace{-8mm}
\centerline{(a) \hspace{7cm} (b)}
\centerline{\includegraphics[width=8cm] {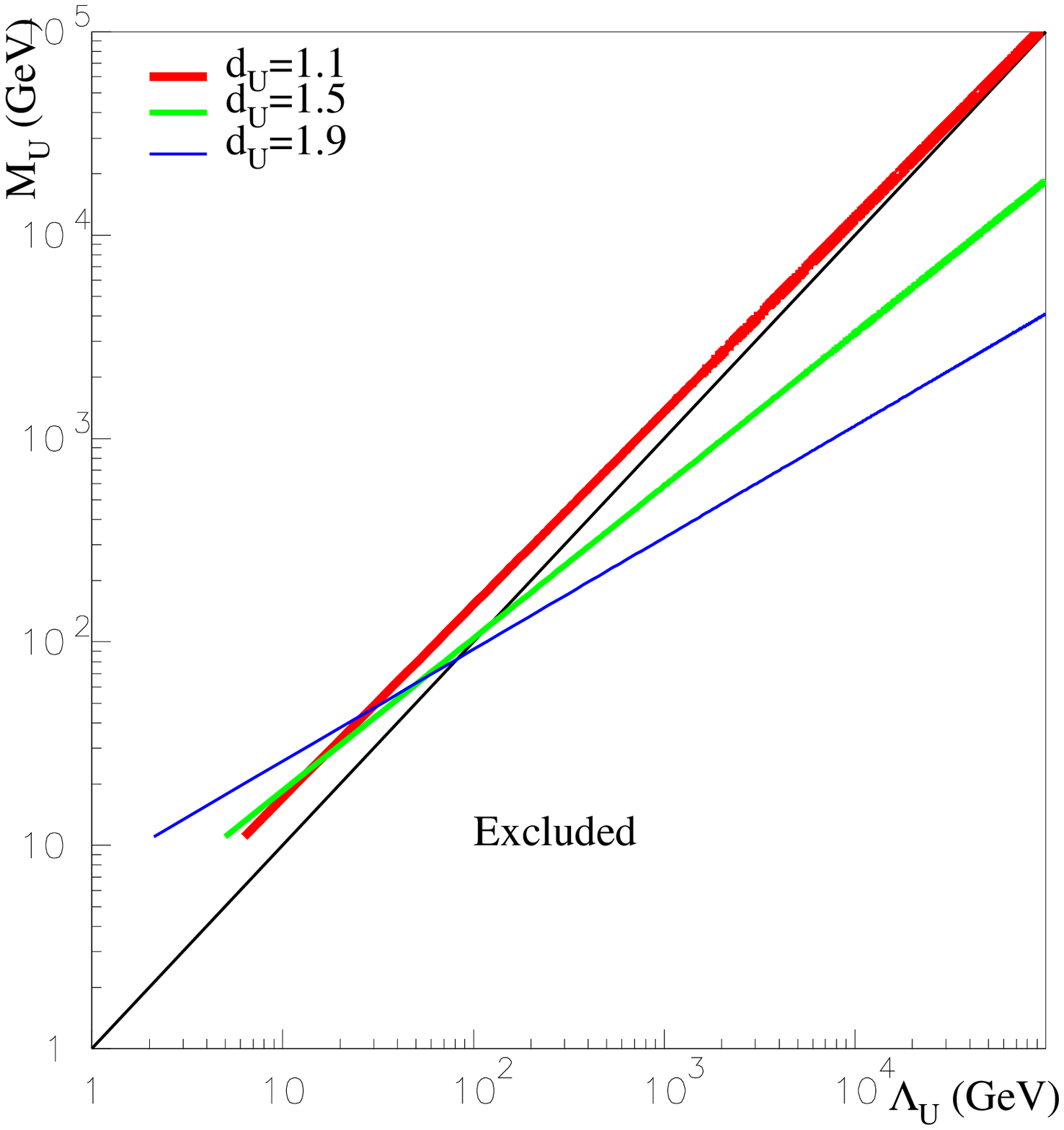}
\includegraphics[width=8cm] {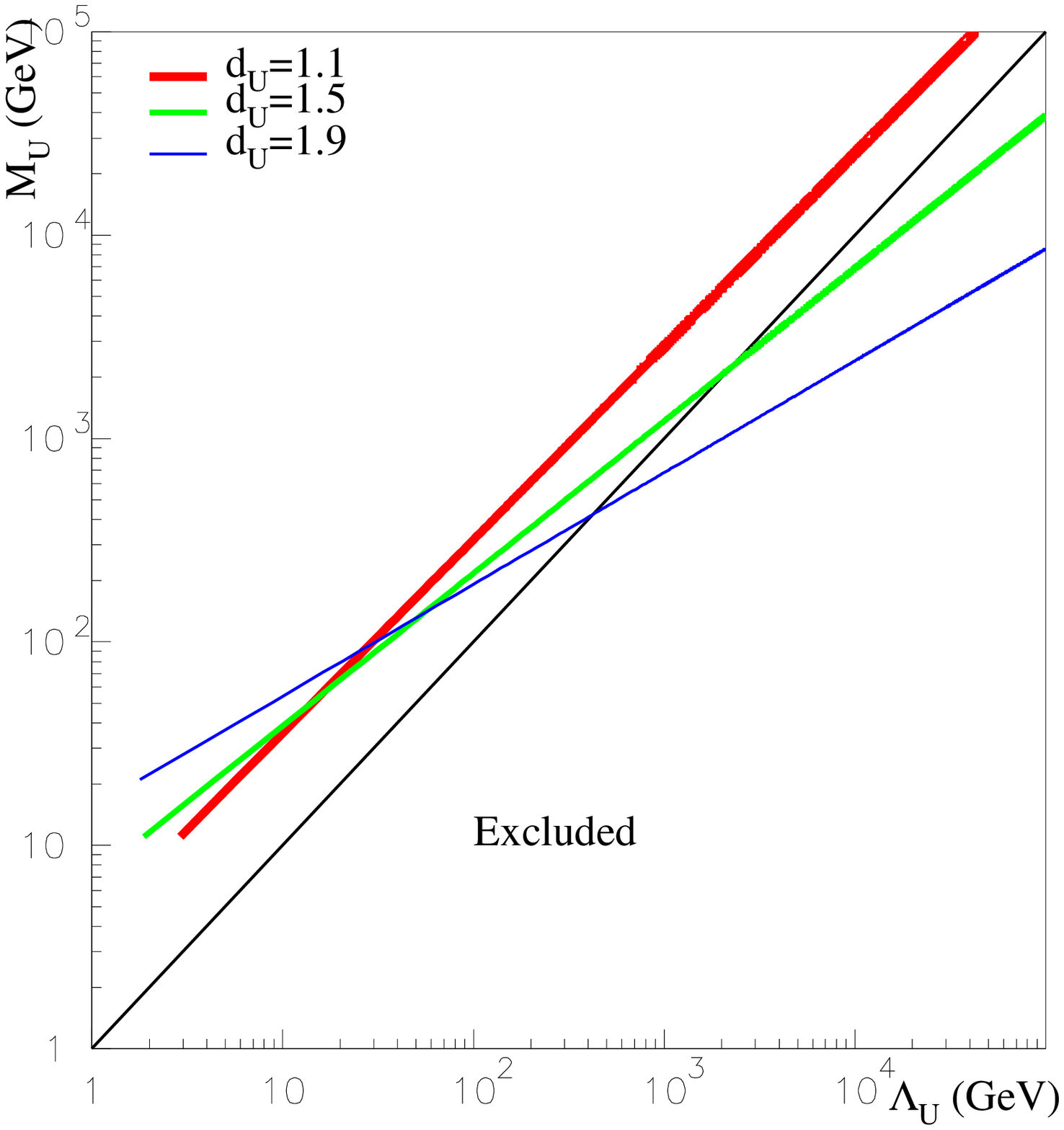}}
\vspace{-8mm}
\centerline{(c) \hspace{7cm} (d)}
\caption{ \label{sparameter}
The contour plot of the $S$ parameter within $1\sigma$ bound
in $\Lambda_{\cal U}$ and $M_{\cal U}$ plane
for $d_{\cal U}=1.1$ (red), $d_{\cal U}=1.5$ (green),
and $d_{\cal U}=1.9$ (blue) with $\mu=1$ GeV in (a), (b)
and $\mu=m_Z$ in (c), (d).
$\lambda_b=-\lambda_w=1$ in case (a) and (c),
$\lambda_b=\lambda_w=1$ in case (b) and (d).
The region above the corresponding curve for different $d_{\cal U}$
is allowed with $1\sigma$ bound.
}
\end{figure}

The overall $S$ is the sum of the unparticle and the SM contributions, 
where we assume a light SM Higgs $m_H = 120$ GeV, and $S_{\rm SM} = - 0.22$.
From the electroweak precision data, the S is determined as 
$S = -0.13 \pm 0.10$~\cite{Yao:2006px}.
In our numerical results,
we choose dimensionless couplings to $1$,
$d_{\rm UV}=3$ for $d_{\cal U}=1.1$, $1.5$, $1.9$,
and choose $\mu=1$ GeV and $\mu=m_Z$.
We plot the contour diagram of the $S$ parameter with
$1\sigma$ bound in $( \Lambda_{\cal U} , M_{\cal U} )$ plane
for different $d_{\cal U}$'s specified in the caption
in Fig.~\ref{sparameter} (a), (c) with different sign of $\lambda_b$ and 
$\lambda_w$ and Fig.~\ref{sparameter} (b), (d) with the same signs.
In Fig.~\ref{sparameter} (a), (c),
the contribution from unparticle is positive,
and the curve in the plot corresponds to $1\sigma$ upper bound $S=-0.03$,
while the curve in Fig.(\ref{sparameter}b) corresponds to
$1\sigma$ lower bound $S=-0.23$, and unparticle contribution is negative,
and hence constructive with the SM contribution.
The region above the curve is allowed by the $1\sigma$ bound of 
the $S$ parameter.

\subsection{Muon anomalous magnetic dipole moment $(g-2)_\mu$}

The muon anomalous magnetic dipole moment $(g-2)$ is a testing 
ground of the SM at quantum levels, and
also a sensitive probe to new physics scenarios at EW scales. 
At present, the experimental data and the SM prediction  
have $3.4\sigma$ deviation~\cite{Miller:2007kk}, 
\[
a^{\rm exp}_\mu - a^{\rm SM}_\mu = (29.5 \pm 8.8) \times 10^{-10}.
\]

The antisymmetric rank--2 tensor unparticle operator can contribute to
the muon $(g-2)_\mu$ via  both 
$\lambda_b - \lambda_f$ and $\lambda_f - \lambda_w$ interactions. 
We can derive the contribution $a_\mu \equiv \frac{1}{2} (g - 2)_\mu$ as
\begin{eqnarray}
\label{g-2}
a^{\cal U}_\mu = \lambda_b \lambda_f
  \frac{A_{d_{\cal U}}}{2 \sin(d_{\cal U}\pi)}
  \frac{m_\mu^2 (\mu^2)^{d_{\cal U} -2}}{\Lambda_2^{d_{\cal U}-2}
   \Lambda_4^{d_{\cal U}}}
  \left(1 - \frac{s_w}{4 c_w} \frac{g}{g^\prime} \frac{\lambda_w}{\lambda_b}
   \frac{v^2}{\Lambda_4^2}
    \frac{\Lambda_2^{d_{\cal U}-2}}{\Lambda_4^{d_{\cal U}-2}} \right) \,.
\end{eqnarray}
Note that this is divergent for $\mu = 0$ and $1 < d_{\cal U} < 2$, and
we choose a nonzero $\mu$ in the numerical analysis: $\mu = 1$ GeV and
$\mu = m_Z$. 

\begin{figure}
\centerline{\includegraphics[width=8cm] {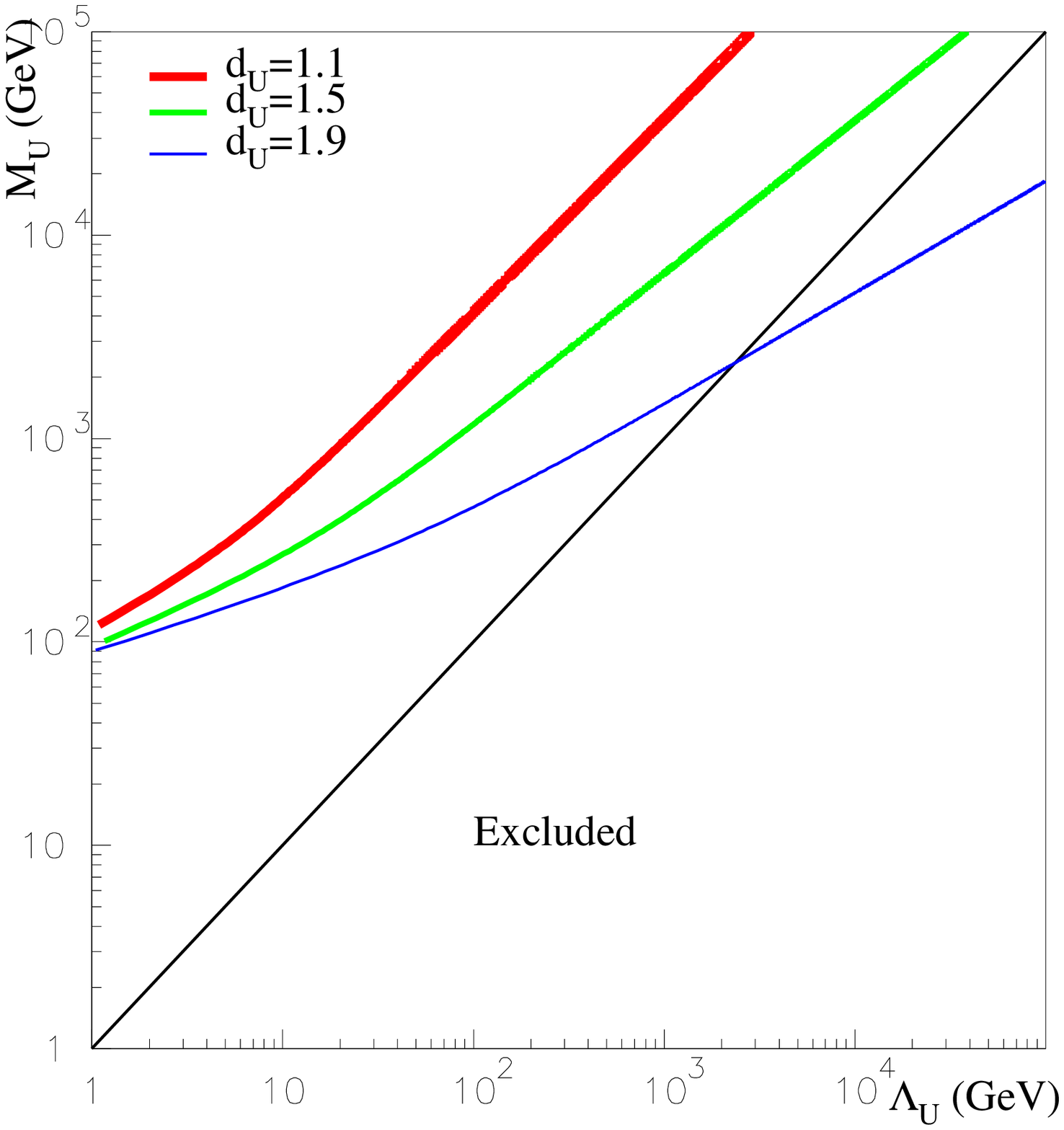}
\includegraphics[width=8cm] {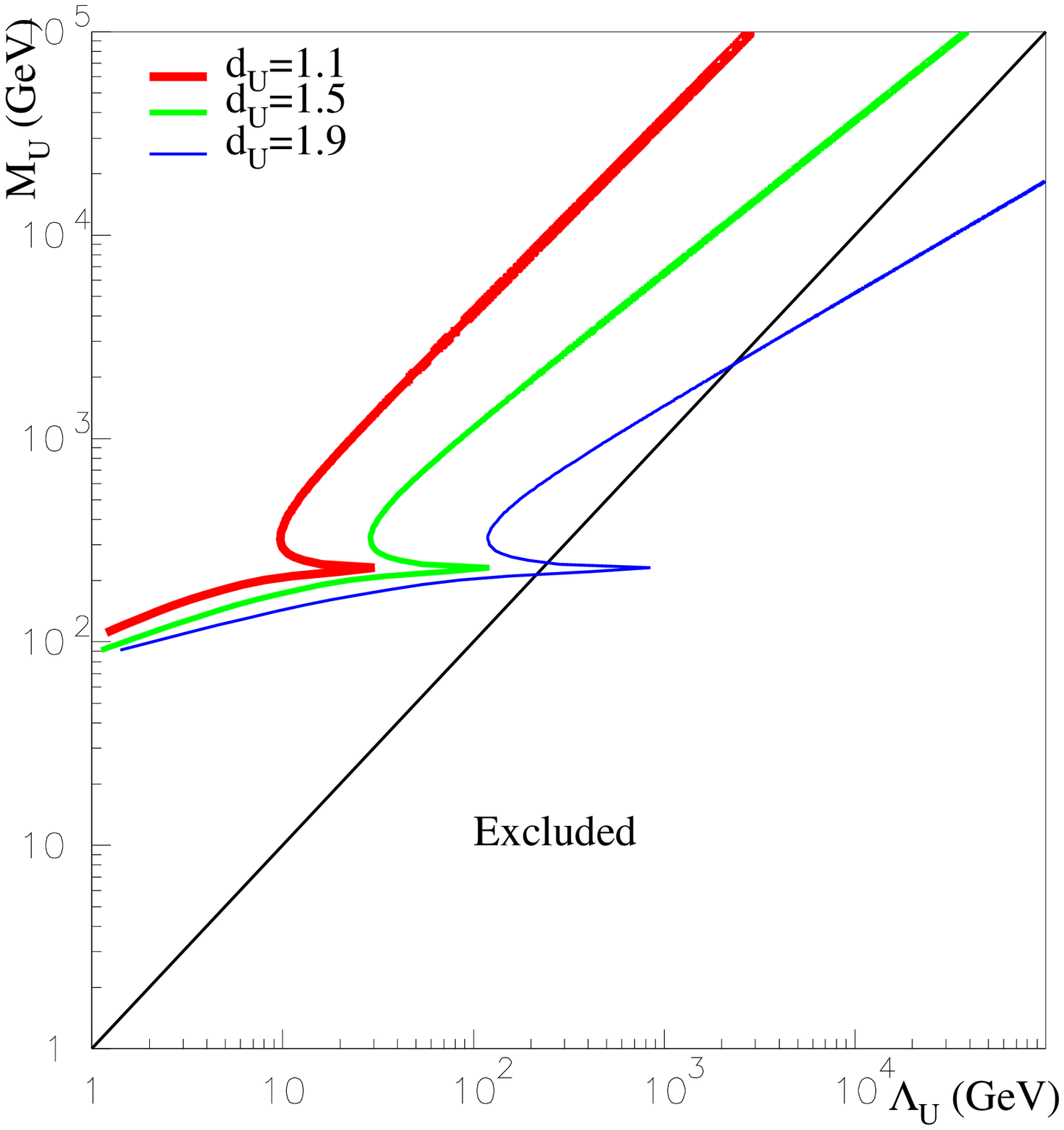}}
\vspace{-8mm}
\centerline{(a) \hspace{7cm} (b)}
\centerline{\includegraphics[width=8cm] {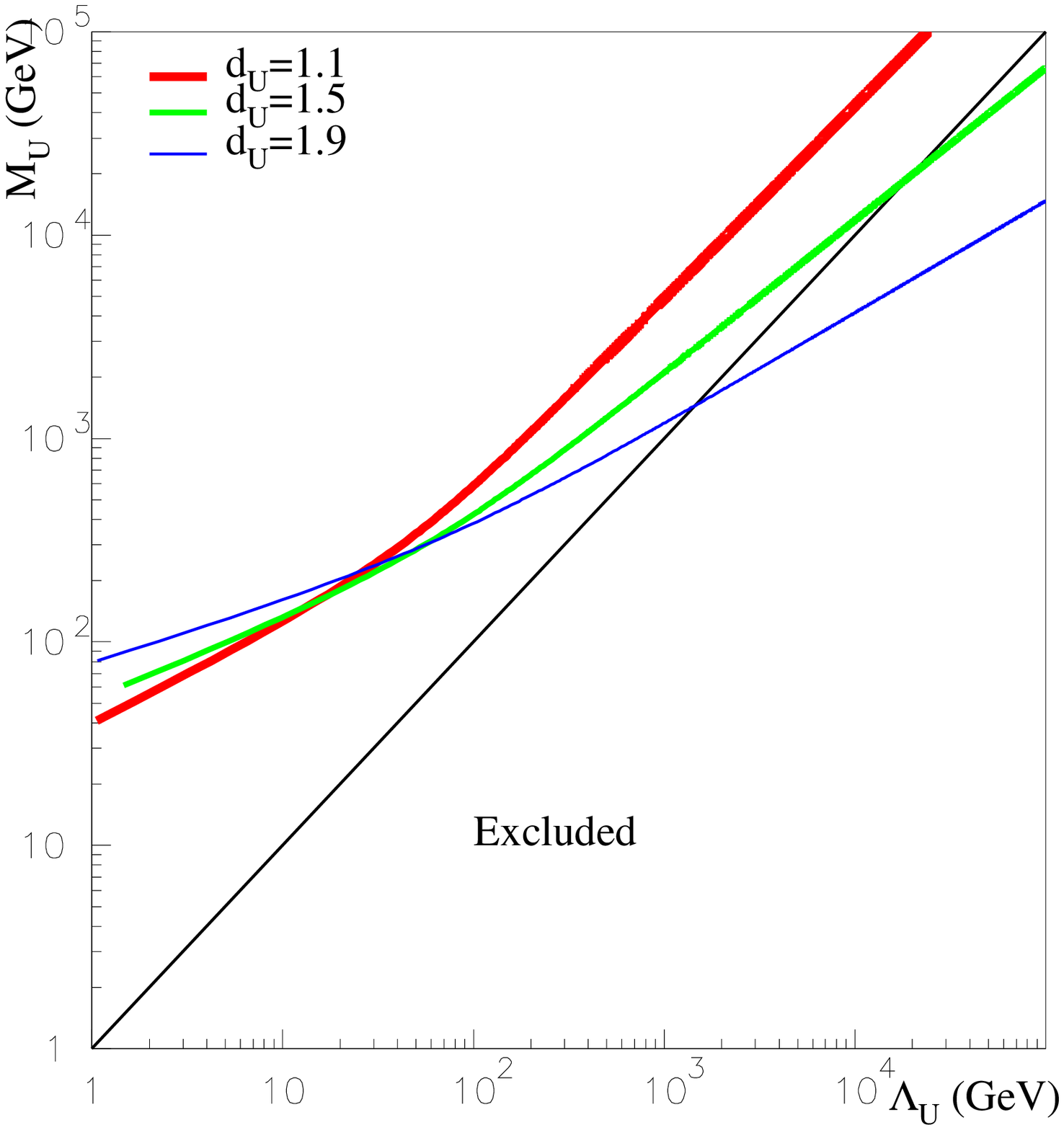}
\includegraphics[width=8cm] {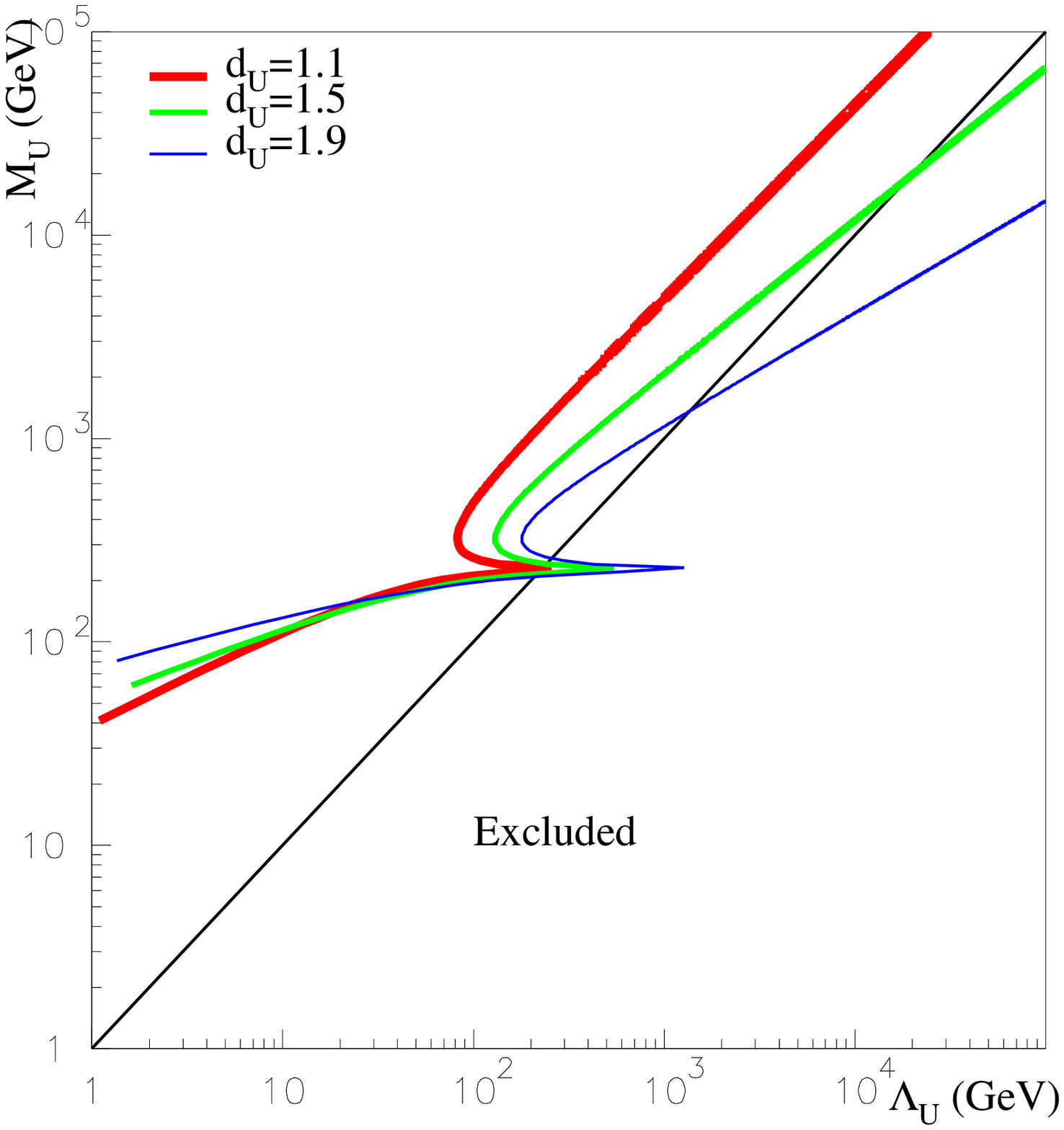}}
\vspace{-8mm}
\centerline{(c) \hspace{7cm} (d)}
\caption{ \label{muon}
The contour plot of $a_\mu$ with $a^{\cal U}_\mu = 10^{-9}$
in $\Lambda_{\cal U}$ and $M_{\cal U}$ plane
for $d_{\cal U}=1.1$ (red), $d_{\cal U}=1.5$ (green),
and $d_{\cal U}=1.9$ (blue) with $\mu=1$ GeV in (a), (b)
and $\mu=m_Z$ in (c), (d).
$-\lambda_f=\lambda_b=-\lambda_w=1$ in case (a) and (c),
$-\lambda_f=\lambda_b=\lambda_w=1$ in case (b) and (d).
In the region above the corresponding curve for different $d_{\cal U}$,
$a^{\cal U}_\mu$ is smaller than $10^{-9}$.
}
\end{figure}

In the numerical calculation, 
we fix the dimensionless couplings to $1$
$d_{\rm UV}=3$ for $d_{\cal U}=1.1$, $1.5$, $1.9$.
There could be additional contributions to the muon $(g-2)$ from 
other unparticle operators. 
Therefore, instead of fitting the muon $(g-2)$ by the antisymmetric 
rank--2 tensor unparticle operator, we assume its contribution is smaller
than $\sim 10^{-9}$.  In Fig.~\ref{muon},  we show the contour plots of 
unparticle induced $a_\mu$ with $a^{\cal U}_\mu=10^{-9}$ 
in the $( \Lambda_{\cal U} , M_{\cal U} )$ plane. 
We find that the contribution from unparticle can be as large as $10^{-9}$ 
for $d_{\cal U} = 1.1, 1.5, 1.9$ in the region below the curves, 
which are disfavored accordingly. Note that the constraint from the muon
$(g-2)$ is stronger than those obtained from LEP/SLC data 
\cite{Bander:2007nd}. 

\section{Conclusions}
In this paper, we concentrated on the antisymmetric rank--2 tensor
unparticle operator ${\cal O}_{\cal U,A}^{\mu\nu}$ with scaling dimension 
$1 < d_{\cal U} <2$. This operator has a unique property that 
it can mix with the $B_{\mu\nu}$ operator of the SM sector,
and can modify the properties of $Z^0$ bosons.
We first derived the two-point function and the propagator for the
antisymmetric rank--2 tensor unparticle operator, studied their effects
on $Z^0$ boson properties, the invisible $Z^0$ decay width 
$Z^0 \rightarrow {\cal U}$, $Z^0 \rightarrow b\bar{b}$, the $S$ parameter 
and the muon $(g-2)_\mu$. 
We find that the last two observables are divergent for 
$1 < d_{\cal U} <2$. Therefore we had to introduce a phenomenological 
parameter $\mu$, a low energy scale where scale symmetry is broken, 
in order to make $S$ parameter and the muon $(g-2)$ finite.
The most stringent bounds on the fundamental scales $\Lambda_{\cal U}$ and
$M_{\cal U}$ come from the invisible decay width $Z^0 \rightarrow {\cal U}$
for $\mu < m_Z$. We find  $r = \Lambda_{\cal U}/M_{\cal U} 
\lesssim 0.1$ is favored, which is more stringent than the bounds from 
LEP/SLC \cite{Bander:2007nd}.  
Furthermore the contributions of ${\cal O}_{\cal U,A}^{\mu\nu}$ to the 
$S$ parameter and the muon $(g-2)$ are proportional to 
$( \mu^2 )^{d_{\cal U}- 2}$, which is divergent for $\mu^2 = 0$ 
for $1 < d_{\cal U} < 2$. Therefore we need nonzero $\mu^2$ 
for $1 < d_{\cal U} < 2$. For $\mu \neq 0$ (especially for $\mu \geq m_Z$), 
there would be no constraint 
from the invisible $Z^0$ decay width or from astrophysical processes 
involving unparticle emissions from stars and supernovae.  
Still the constraint from the muon $(g-2)_\mu$ is quite significant for 
$\mu \neq 0$.

\acknowledgements
This work is supported in part by KOSEF through CHEP at 
Kyungpook National University.



\end{document}